\documentclass[italian,english]{article}
\usepackage[latin9]{inputenc}
\usepackage{textcomp}
\usepackage{color}
\usepackage{graphicx}
\usepackage{amssymb}

\makeatletter
\newcommand{\lyxaddress}[1]{
\par {\raggedright #1
\vspace{1.4em}
\noindent\par}
}

\makeatother

\usepackage{babel}

\begin{document}

\title{\textbf{Massive gravitational waves from $f(R)$ theories of gravity:
potential detection with LISA}}

\author{\textbf{\textonesuperior{}Salvatore Capozziello, \texttwosuperior{}Christian
Corda and \textthreesuperior{}Maria Felicia De Laurentis}}

\maketitle
\begin{center}
\textonesuperior{}Dipartimento di Scienze Fisiche, Università di Napoli
{}``Federico II'', INFN Sezione di Napoli, Compl. Univ di Monte
S. Angelo, Edificio G, via Cinthia, I-80126, Napoli, Italy
\par\end{center}

\texttwosuperior{}Associazione Galileo Galilei, Via Pier Cironi 16
- 59100 PRATO, Italy and Centro di Scienze Naturali, Via di Galceti
74 - 59100 PRATO, Italy

\begin{center}
\textthreesuperior{} Relativity and Gravitation Group Politecnico
di Torino, Corso Duca degli Abruzzi 24, I - 10129 Torino, Italy
\par\end{center}

\lyxaddress{\begin{center}
\textit{E-mail addresses:} \textcolor{blue}{\textonesuperior{}capozzie@na.infn.it;
\texttwosuperior{}christian.corda@ego-gw.it; \textthreesuperior{}
mariafelicia.delaurentist@polito.it}
\par\end{center}}

\begin{abstract}
This letter is a generalization of previous results on gravitational
waves (GWs) from $f(R)$ theories of gravity. In some previous papers,
particular $f(R$) theories have been linearized for the first time
in the literature. Now, the process is further generalized, showing
that every $f(R$) theory can be linearized producting a third massive
mode of gravitational radiation. In this framework, previous results
are particular cases of the more general problem that is discussed
in this letter. The potential detectability of such massive GWs with
LISA is also discussed with the auxilium of longitudinal response
functions.
\end{abstract}
Recently, the data analysis of interferometric GWs detectors has been
started (for the current status of GWs interferometers see \cite{key-1,key-2,key-3,key-4,key-5,key-6,key-7,key-8})
and the scientific community aims at a first direct detection of GWs
in next years. 

Detectors for GWs will be important for a better knowledge of the
Universe and either to confirm or ruling out the physical consistency
of General Relativity or any other theory of gravitation \cite{key-9,key-10,key-11,key-12,key-13,key-14}. 

An early, but quit interesting, work on GWs from other theories of
gravity has been wriiten by Wagoner 

In this letter, the production of a hypotetical massive component
of gravitational radiation which arises from a general $f(R)$ theory
of gravity is shown. The presence of the mass could also have important
applications in cosmology because the fact that GWs can have mass
could give a contribution to the dark matter of the Universe. 

The first and simplest $f(R)$ theory of gravity was proposed by Starobinsky
\cite{key-15}, who dicussed the action \begin{equation}
S=\int d^{4}x\sqrt{-g}(R+\alpha R^{2})+\mathcal{L}_{m}.\label{eq: high order 1}\end{equation}
The production and the potential detection of GWs from this theory
has been analysed in \cite{key-16}. In \cite{key-17}, it has been
also shown that, from this particular linearized theory, it is possible
to obtain an oscillating model of Universe.

Another example is the action \begin{equation}
S=\int d^{4}x\sqrt{-g}(R+R^{-1})+\mathcal{L}_{m}.\label{eq: high order 1 bis}\end{equation}

This action has been analysed in a cosmological context in \cite{key-18},
while the GWs case has been analysed in \cite{key-19}. This kind
of theory could, in principle, be connected with the Dark Matter problem
too\cite{key-20}. Criticisms on $f(R)$ theories of gravity arises
from the fact that lots of such theories can be excluded by requirements
of Cosmology and Solar System tests \cite{key-24,key-25}. It is important
to emphasize that the theory of eq. (\ref{eq: high order 1 bis}),
differently from the theory of the action (\ref{eq: high order 1})
is not in conflict with such constrains \cite{key-24,key-25}. In
this context, even the theory 

\begin{equation}
S=\int d^{4}x\sqrt{-g}f_{0}R^{1+\varepsilon}+\mathcal{L}_{m},\label{eq: high order 1 tris}\end{equation}

which has been discussed in \cite{key-25}, is quite interesting.

Equations (\ref{eq: high order 1}), (\ref{eq: high order 1 bis})
and (\ref{eq: high order 1 tris}) are particular choices in respect
to the well known canonical one of general relativity (the Einstein
- Hilbert action \cite{key-21,key-22}) which is 

\begin{equation}
S=\int d^{4}x\sqrt{-g}R+\mathcal{L}_{m},\label{eq: EH}\end{equation}

where $R$ is the Ricci scalar curvature. 

Now, we will analyse the general case, i.e. \begin{equation}
S=\int d^{4}x\sqrt{-g}f(R)+\mathcal{L}_{m},\label{eq: high order generale}\end{equation}

where $f(R)$ is a generic high order theory of gravity.

Of course, the cases which have been analysed in \cite{key-16} and
in \cite{key-19} are particular cases of the more general case that
we are going to analyse now. 

As we will interact with gravitational waves, i.e. the linearized
theory in vacuum, $\mathcal{L}_{m}=0$ will be put and the pure curvature
action \begin{equation}
S=\int d^{4}x\sqrt{-g}f(R)\label{eq: high order 12}\end{equation}
 will be considered. 

By varying the action (\ref{eq: high order 12}) in respect to $g_{\mu\nu}$
(see refs. \cite{key-16,key-17,key-19} for a parallel computation)
the field equations are obtained (note that in this paper we work
with $G=1$, $c=1$ and $\hbar=1$):

\begin{equation}
f'(R)R_{\mu\nu}-\frac{1}{2}f(R)g_{\mu\nu}-f'(R)_{;\mu;\nu}+g_{\mu\nu}\square f'(R)=0\label{eq: einstein-general}\end{equation}

which are the modified Einstein field equations. $f'(R)$ is the derivative
of $f$ in respect to the Ricci scalar. Writing down, exlplicitly,
the Einstein tensor, eqs. (\ref{eq: einstein-general}) become\begin{equation}
G_{\mu\nu}=\frac{1}{f'(R)}\{\frac{1}{2}g_{\mu\nu}[f(R)-f'(R)R]+f'(R)_{;\mu;\nu}-g_{\mu\nu}\square f'(R)\}.\label{eq: einstein 2}\end{equation}

Taking the trace of the field equations (\ref{eq: einstein 2}) one
gets 

\begin{equation}
3\square f'(R)+Rf'(R)-2f(R)=0,\label{eq: KG}\end{equation}

and, with the identifications \cite{key-23}

\begin{equation}
\begin{array}{ccccc}
\Phi\rightarrow f'(R) &  & \textrm{and } &  & \frac{dV}{d\Phi}\rightarrow\frac{2f(R)-Rf'(R)}{3}\end{array}\label{eq: identifica}\end{equation}

a Klein - Gordon equation for the effective $\Phi$ scalar field is
obtained:

\begin{equation}
\square\Phi=\frac{dV}{d\Phi}.\label{eq: KG2}\end{equation}

To study gravitational waves, the linearized theory has to be analyzed,
with a little perturbation of the background, which is assumed given
by a a Minkowskian background plus $\Phi=\Phi_{0},$i.e. we are linearizing
into a background with constant curvature \cite{key-19,key-24}. We
also assume $\Phi_{0}$ to be a minimum for $V$: 

\begin{equation}
V\simeq\frac{1}{2}\alpha\delta\Phi^{2}\Rightarrow\frac{dV}{d\Phi}\simeq m^{2}\delta\Phi,\label{eq: minimo}\end{equation}

and the constant $m$ has mass dimension. 

Putting

\begin{equation}
\begin{array}{c}
g_{\mu\nu}=\eta_{\mu\nu}+h_{\mu\nu}\\
\\\Phi=\Phi_{0}+\delta\Phi.\end{array}\label{eq: linearizza}\end{equation}

to first order in $h_{\mu\nu}$ and $\delta\Phi$, calling $\widetilde{R}_{\mu\nu\rho\sigma}$
, $\widetilde{R}_{\mu\nu}$ and $\widetilde{R}$ the linearized quantity
which correspond to $R_{\mu\nu\rho\sigma}$ , $R_{\mu\nu}$ and $R$,
the linearized field equations are obtained \cite{key-16,key-17,key-19}:

\begin{equation}
\begin{array}{c}
\widetilde{R}_{\mu\nu}-\frac{\widetilde{R}}{2}\eta_{\mu\nu}=(\partial_{\mu}\partial_{\nu}h_{f}-\eta_{\mu\nu}\square h_{f})\\
\\{}\square h_{f}=m^{2}h_{f},\end{array}\label{eq: linearizzate1}\end{equation}

where 

\begin{equation}
h_{f}\equiv\frac{\delta\Phi}{\Phi_{0}}.\label{eq: definizione}\end{equation}

$\widetilde{R}_{\mu\nu\rho\sigma}$ and eqs. (\ref{eq: linearizzate1})
are invariants for gauge transformations \cite{key-16,key-17,key-19}

\begin{equation}
\begin{array}{c}
h_{\mu\nu}\rightarrow h'_{\mu\nu}=h_{\mu\nu}-\partial_{(\mu}\epsilon_{\nu)}\\
\\\delta\Phi\rightarrow\delta\Phi'=\delta\Phi;\end{array}\label{eq: gauge}\end{equation}

then 

\begin{equation}
\bar{h}_{\mu\nu}\equiv h_{\mu\nu}-\frac{h}{2}\eta_{\mu\nu}+\eta_{\mu\nu}h_{f}\label{eq: ridefiniz}\end{equation}

can be defined, and, considering the transformation for the parameter
$\epsilon^{\mu}$

\begin{equation}
\square\epsilon_{\nu}=\partial^{\mu}\bar{h}_{\mu\nu},\label{eq:lorentziana}\end{equation}
 a gauge parallel to the Lorenz one of electromagnetic waves can be
choosen:

\begin{equation}
\partial^{\mu}\bar{h}_{\mu\nu}=0.\label{eq: cond lorentz}\end{equation}

In this way, field equations read like

\begin{equation}
\square\bar{h}_{\mu\nu}=0\label{eq: onda T}\end{equation}

\begin{equation}
\square h_{f}=m^{2}h_{f}\label{eq: onda S}\end{equation}

Solutions of eqs. (\ref{eq: onda T}) and (\ref{eq: onda S}) are
plan waves \cite{key-16,key-17,key-19}:

\begin{equation}
\bar{h}_{\mu\nu}=A_{\mu\nu}(\overrightarrow{p})\exp(ip^{\alpha}x_{\alpha})+c.c.\label{eq: sol T}\end{equation}

\begin{equation}
h_{f}=a(\overrightarrow{p})\exp(iq^{\alpha}x_{\alpha})+c.c.\label{eq: sol S}\end{equation}

where

\begin{equation}
\begin{array}{ccc}
k^{\alpha}\equiv(\omega,\overrightarrow{p}) &  & \omega=p\equiv|\overrightarrow{p}|\\
\\q^{\alpha}\equiv(\omega_{m},\overrightarrow{p}) &  & \omega_{m}=\sqrt{m^{2}+p^{2}}.\end{array}\label{eq: k e q}\end{equation}

In eqs. (\ref{eq: onda T}) and (\ref{eq: sol T}) the equation and
the solution for the standard waves of General Relativity \cite{key-21,key-22}
have been obtained, while eqs. (\ref{eq: onda S}) and (\ref{eq: sol S})
are respectively the equation and the solution for the massive mode
(see also \cite{key-16,key-17,key-19}).

The fact that the dispersion law for the modes of the massive field
$h_{f}$ is not linear has to be emphatized. The velocity of every
{}``ordinary'' (i.e. which arises from General Relativity) mode
$\bar{h}_{\mu\nu}$ is the light speed $c$, but the dispersion law
(the second of eq. (\ref{eq: k e q})) for the modes of $h_{f}$ is
that of a massive field which can be discussed like a wave-packet
\cite{key-16,key-17,key-19}. Also, the group-velocity of a wave-packet
of $h_{f}$ centered in $\overrightarrow{p}$ is 

\begin{equation}
\overrightarrow{v_{G}}=\frac{\overrightarrow{p}}{\omega},\label{eq: velocita' di gruppo}\end{equation}

which is exactly the velocity of a massive particle with mass $m$
and momentum $\overrightarrow{p}$.

From the second of eqs. (\ref{eq: k e q}) and eq. (\ref{eq: velocita' di gruppo})
it is simple to obtain:

\begin{equation}
v_{G}=\frac{\sqrt{\omega^{2}-m^{2}}}{\omega}.\label{eq: velocita' di gruppo 2}\end{equation}

Then, wanting a constant speed of the wave-packet, it has to be \cite{key-16,key-17,key-19}

\begin{equation}
m=\sqrt{(1-v_{G}^{2})}\omega.\label{eq: relazione massa-frequenza}\end{equation}

Now, the analysis can remain in the Lorenz gauge with trasformations
of the type $\square\epsilon_{\nu}=0$; this gauge gives a condition
of transversality for the ordinary part of the field: $k^{\mu}A_{\mu\nu}=0$,
but does not give the transversality for the total field $h_{\mu\nu}$.
From eq. (\ref{eq: ridefiniz}) it is

\begin{equation}
h_{\mu\nu}=\bar{h}_{\mu\nu}-\frac{\bar{h}}{2}\eta_{\mu\nu}+\eta_{\mu\nu}h_{f}.\label{eq: ridefiniz 2}\end{equation}

At this point, if being in the massless case \cite{key-16,key-17,key-19},
it could been put

\begin{equation}
\begin{array}{c}
\square\epsilon^{\mu}=0\\
\\\partial_{\mu}\epsilon^{\mu}=-\frac{\bar{h}}{2}+h_{f},\end{array}\label{eq: gauge2}\end{equation}

which gives the total transversality of the field. But in the massive
case this is impossible. In fact, applying the Dalembertian operator
to the second of eqs. (\ref{eq: gauge2}) and using the field equations
(\ref{eq: onda T}) and (\ref{eq: onda S}) it results

\begin{equation}
\square\epsilon^{\mu}=m^{2}h_{f},\label{eq: contrasto}\end{equation}

which is in contrast with the first of eqs. (\ref{eq: gauge2}). In
the same way, it is possible to show that it does not exist any linear
relation between the tensorial field $\bar{h}_{\mu\nu}$ and the massive
field $h_{f}$. Thus a gauge in wich $h_{\mu\nu}$ is purely spatial
cannot be chosen (i.e. it cannot be put $h_{\mu0}=0,$ see eq. (\ref{eq: ridefiniz 2}))
. But the traceless condition to the field $\bar{h}_{\mu\nu}$ can
be put :

\begin{equation}
\begin{array}{c}
\square\epsilon^{\mu}=0\\
\\\partial_{\mu}\epsilon^{\mu}=-\frac{\bar{h}}{2}.\end{array}\label{eq: gauge traceless}\end{equation}

These equations imply

\begin{equation}
\partial^{\mu}\bar{h}_{\mu\nu}=0.\label{eq: vincolo}\end{equation}

To save the conditions $\partial_{\mu}\bar{h}^{\mu\nu}$ and $\bar{h}=0$
transformations like

\begin{equation}
\begin{array}{c}
\square\epsilon^{\mu}=0\\
\\\partial_{\mu}\epsilon^{\mu}=0\end{array}\label{eq: gauge 3}\end{equation}

can be used and, taking $\overrightarrow{p}$ in the $z$ direction,
a gauge in which only $A_{11}$, $A_{22}$, and $A_{12}=A_{21}$ are
different to zero can be chosen. The condition $\bar{h}=0$ gives
$A_{11}=-A_{22}$. Now, putting these equations in eq. (\ref{eq: ridefiniz 2}),
it results

\begin{equation}
h_{\mu\nu}(t,z)=A^{+}(t-z)e_{\mu\nu}^{(+)}+A^{\times}(t-z)e_{\mu\nu}^{(\times)}+h_{f}(t-v_{G}z)\eta_{\mu\nu}.\label{eq: perturbazione totale}\end{equation}

The term $A^{+}(t-z)e_{\mu\nu}^{(+)}+A^{\times}(t-z)e_{\mu\nu}^{(\times)}$
describes the two standard polarizations of gravitational waves which
arise from General Relativity, while the term $h_{f}(t-v_{G}z)\eta_{\mu\nu}$
is the massive field arising from the generic high order $f(R$) theory.
In other words, the function $f'(R$) of the Ricci scalar generates
a third massive polarization for gravitational waves which is not
present in standard General Relativity. Note that the line element
(\ref{eq: perturbazione totale}) has been obtained in both of references
\cite{key-16} and \cite{key-19} starting by the actions (\ref{eq: high order 1})
and (\ref{eq: high order 1 bis}) respectively. Here we have shown
that such a line element, i.e. the presence of a third massive polarization,
is proper to \textbf{every} $f(R)$ theory of gravity.

The analysis of the two standard polarization is well known in the
literature \cite{key-2,key-3,key-21,key-22}. For a the pure polarization
arising from the $f(R$) theory eq. (\ref{eq: perturbazione totale})
can be rewritten as

\begin{equation}
h_{\mu\nu}(t-v_{G}z)=h_{f}(t-v_{G}z)\eta_{\mu\nu}\label{eq: perturbazione scalare}\end{equation}
and the corrispondent line element is the conformally flat one

\begin{equation}
ds^{2}=[1+h_{f}(t-v_{G}z)](-dt^{2}+dz^{2}+dx^{2}+dy^{2}).\label{eq: metrica puramente scalare}\end{equation}

In \cite{key-19} it has been shown that in this kind of line element
the effect of the mass is the generation of a \textit{longitudinal}
force (in addition to the transverse one) while in the limit $m\rightarrow0$
the longitudinal force vanishes. 

Now, before starting the analysis, it has to be discussed if there
are fenomenogical limitations to the mass of the GW \cite{key-19,key-24}.
A strong limitation arises from the fact that the GW needs a frequency
which falls in the frequency-range for both of earth based and space
based gravitational antennas, that is the interval $10^{-4}Hz\leq f\leq10KHz$
\cite{key-1,key-4,key-5,key-6,key-7,key-8,key-26,key-27}. For a massive
GW, from \cite{key-12,key-14,key-16,key-19} it is:

\begin{equation}
2\pi f=\omega=\sqrt{m^{2}+p^{2}},\label{eq: frequenza-massa}\end{equation}

were $p$ is the momentum. Thus, it needs

\begin{equation}
0eV\leq m\leq10^{-11}eV.\label{eq: range di massa}\end{equation}

A stronger limitation is given by requirements of cosmology and Solar
System tests on extended theories of gravity. In this case it is \cite{key-24} 

\begin{equation}
0eV\leq m\leq10^{-33}eV.\label{eq: range di massa 2}\end{equation}

For these light scalars, their effect can be still discussed as a
coherent GW. 

The frequency-dependent response function, for a massive mode of gravitational
radiation, has been obtained in \cite{key-19} for the particular
case $f(R)=R+R^{-1}.$ Here, the computation will be performed with
another treatment, showing that the longitudinal response function
which has been found in \cite{key-19} is proper to \textbf{every}
$f(R)$ theory of gravity, and the results will be applied to LISA,
following the advice in \cite{key-24}.

Eq. (\ref{eq: metrica puramente scalare}) can be rewritten as 

\begin{equation}
(\frac{dt}{d\tau})^{2}-(\frac{dx}{d\tau})^{2}-(\frac{dy}{d\tau})^{2}-(\frac{dz}{d\tau})^{2}=\frac{1}{(1+h_{f})},\label{eq: Sh2}\end{equation}

where $\tau$ is the proper time of the test masses.

From eqs. (\ref{eq: metrica puramente scalare}) and (\ref{eq: Sh2})
the geodesic equations of motion for test masses (i.e. the beam-splitter
and the mirrors of the interferometer), can be obtained\begin{equation}
\begin{array}{ccc}
\frac{d^{2}x}{d\tau^{2}} & = & 0\\
\\\frac{d^{2}y}{d\tau^{2}} & = & 0\\
\\\frac{d^{2}t}{d\tau^{2}} & = & \frac{1}{2}\frac{\partial_{t}(1+h_{f})}{(1+h_{f})^{2}}\\
\\\frac{d^{2}z}{d\tau^{2}} & = & -\frac{1}{2}\frac{\partial_{z}(1+h_{f})}{(1+h_{f})^{2}}.\end{array}\label{eq: geodetiche Corda}\end{equation}

The first and the second of eqs. (\ref{eq: geodetiche Corda}) can
be immediately integrated obtaining

\begin{equation}
\frac{dx}{d\tau}=C_{1}=const.\label{eq: integrazione x}\end{equation}

\begin{equation}
\frac{dy}{d\tau}=C_{2}=const.\label{eq: integrazione x}\end{equation}

In this way eq. (\ref{eq: Sh2}) becomes\begin{equation}
(\frac{dt}{d\tau})^{2}-(\frac{dz}{d\tau})^{2}=\frac{1}{(1+h_{f})}.\label{eq: Ch3}\end{equation}

If we assume that test masses are at rest initially we get $C_{1}=C_{2}=0$.
Thus we see that, even if the GW arrives at test masses, we do not
have motion of test masses within the $x-y$ plane in this gauge.
We could understand this directly from eq. (\ref{eq: metrica puramente scalare})
because the absence of the $x$ and of the $y$ dependences in the
metric implies that test masses momentum in these directions (i.e.
$C_{1}$ and $C_{2}$ respectively) is conserved. This results, for
example, from the fact that in this case the $x$ and $y$ coordinates
do not esplicitly enter in the Hamilton-Jacobi equation for a test
mass in a gravitational field \cite{key-2}. 

Now we will see that, in presence of the GW, we have motion of test
masses in the $z$ direction which is the direction of the propagating
wave. An analysis of eqs. (\ref{eq: geodetiche Corda}) shows that,
to simplify equations, we can introduce the retarded and advanced
time coordinates ($u,v$):

\begin{equation}
\begin{array}{c}
u=t-v_{G}z\\
\\v=t+v_{G}z.\end{array}\label{eq: ret-adv}\end{equation}

From the third and the fourth of eqs. (\ref{eq: geodetiche Corda})
we have

\begin{equation}
\frac{d}{d\tau}\frac{du}{d\tau}=\frac{\partial_{v}[1+h_{f}(u)]}{(1+h_{f}(u))^{2}}=0.\label{eq: t-z t+z}\end{equation}

This equation can be integrated obtaining

\begin{equation}
\frac{du}{d\tau}=\alpha,\label{eq: t-z}\end{equation}

where $\alpha$ is an integration constant. From eqs. (\ref{eq: Ch3})
and (\ref{eq: t-z}), we also get

\begin{equation}
\frac{dv}{d\tau}=\frac{\beta}{1+h_{f}}\label{eq: t+z}\end{equation}

where $\beta\equiv\frac{1}{\alpha}$, and

\begin{equation}
\tau=\beta u+\gamma,\label{eq: tau}\end{equation}

where the integration constant $\gamma$ correspondes simply to the
retarded time coordinate translation $u$. Thus, without loss of generality,
we can put it equal to zero. Now let us see what is the meaning of
the other integration constant $\beta.$ We can write the equation
for $z$ from eqs. (\ref{eq: t-z}) and (\ref{eq: t+z}):

\begin{equation}
\frac{dz}{d\tau}=\frac{1}{2\beta}(\frac{\beta^{2}}{1+h_{f}}-1).\label{eq: z}\end{equation}

When it is $h_{f}=0$ (i.e. before the GW arrives at the test masses)
eq. (\ref{eq: z}) becomes\begin{equation}
\frac{dz}{d\tau}=\frac{1}{2\beta}(\beta^{2}-1).\label{eq: z ad h nullo}\end{equation}

But this is exactly the initial velocity of the test mass, thus we
have to choose $\beta=1$ because we suppose that test masses are
at rest initially. This also imply $\alpha=1$.

To find the motion of a test mass in the $z$ direction we see that
from eq. (\ref{eq: tau}) we have $d\tau=du$, while from eq. (\ref{eq: t+z})
we have $dv=\frac{d\tau}{1+h_{f}}$. Because it is $v_{G}z=\frac{v-u}{2}$
we obtain

\begin{equation}
dz=\frac{1}{2v_{G}}(\frac{d\tau}{1+h_{f}}-du),\label{eq: dz}\end{equation}

which can be integrated as

\begin{equation}
\begin{array}{c}
z=z_{0}+\frac{1}{2v_{G}}\int(\frac{du}{1+h_{f}}-du)=\\
\\=z_{0}-\frac{1}{2v_{G}}\int_{-\infty}^{t-v_{G}z}\frac{h_{f}(u)}{1+h_{f}(u)}du,\end{array}\label{eq: moto lungo z}\end{equation}

where $z_{0}$ is the initial position of the test mass. Now the displacement
of the test mass in the $z$ direction can be written as

\begin{equation}
\begin{array}{c}
\Delta z=z-z_{0}=-\frac{1}{2v_{G}}\int_{-\infty}^{t-v_{G}z_{0}-v_{G}\Delta z}\frac{h_{f}(u)}{1+h_{f}(u)}du\\
\\\simeq-\frac{1}{2v_{G}}\int_{-\infty}^{t-v_{G}z_{0}}\frac{h_{f}(u)}{1+h_{f}(u)}du.\end{array}\label{eq: spostamento lungo z}\end{equation}
We can also rewrite the results in function of the time coordinate
$t$:

\begin{equation}
\begin{array}{ccc}
x(t) & = & x_{0}\\
\\y(t) & = & y_{0}\\
\\z(t) & = & z_{0}-\frac{1}{2v_{G}}\int_{-\infty}^{t-v_{G}z_{0}}\frac{h_{f}(u)}{1+h_{f}(u)}d(u)\\
\\\tau(t) & = & t-v_{G}z(t),\end{array}\label{eq: moto gauge Corda}\end{equation}

Calling $l$ and $L+l$ the unperturbed positions of the beam-splitter
and of the mirror and using the third of eqs. (\ref{eq: moto gauge Corda})
the varying position of the beam-splitter and of the mirror are given
by

\begin{equation}
\begin{array}{c}
z_{BS}(t)=l-\frac{1}{2v_{G}}\int_{-\infty}^{t-v_{G}l}\frac{h_{f}(u)}{1+h_{f}(u)}d(u)\\
\\z_{M}(t)=L+l-\frac{1}{2v_{G}}\int_{-\infty}^{t-v_{G}(L+l)}\frac{h_{f}(u)}{1+h_{f}(u)}d(u)\end{array}\label{eq: posizioni}\end{equation}

But we are interested in variations in the proper distance (time)
of test masses, thus, in correspondence of eqs. (\ref{eq: posizioni}),
using the fourth of eqs. (\ref{eq: moto gauge Corda}) we get\begin{equation}
\begin{array}{c}
\tau_{BS}(t)=t-v_{G}l-\frac{1}{2}\int_{-\infty}^{t-v_{G}l}\frac{h_{f}(u)}{1+h_{f}(u)}d(u)\\
\\\tau_{M}(t)=t-v_{G}L-v_{G}l-\frac{1}{2}\int_{-\infty}^{t-v_{G}(L+l)}\frac{h_{f}(u)}{1+h_{f}(u)}d(u).\end{array}\label{eq: posizioni 2}\end{equation}

Then the total variation of the proper time is given by

\begin{equation}
\bigtriangleup\tau(t)=\tau_{M}(t)-\tau_{BS}(t)=v_{G}L-\frac{1}{2}\int_{t-v_{G}l}^{t-v_{G}(L+l)}\frac{h_{f}(u)}{1+h_{f}(u)}d(u).\label{eq: time}\end{equation}

In this way, recalling that in the used units the unperturbed proper
distance (time) is $T=L$, the difference between the total variation
of the proper time in presence and the total variation of the proper
time in absence of the GW is \begin{equation}
\delta\tau(t)\equiv\bigtriangleup\tau(t)-L=-L(v_{G}+1)-\frac{1}{2}\int_{t-v_{G}l}^{t-v_{G}(L+l)}\frac{h_{f}(u)}{1+h_{f}(u)}d(u).\label{eq: time variation}\end{equation}

This quantity can be computed in the frequency domain, defining the
Fourier transform of $h_{f}$ as \begin{equation}
\widetilde{h}_{f}(\omega)=\int_{-\infty}^{\infty}dt\textrm{ }h_{f}(t)\exp(i\omega t).\label{eq: trasformata di fourier}\end{equation}

and using the translation and derivation Fourier theorems, obtaining\begin{equation}
\begin{array}{c}
\delta\widetilde{\tau}(\omega)=L(1-v_{G}^{2})\exp[i\omega L(1+v_{G})]+\frac{L}{2\omega L(v_{G}^{2}-1)^{2}}\\
\\{}[\exp[2i\omega L](v_{G}+1)^{3}(-2i+\omega L(v_{G}-1)+2L\exp[i\omega L(1+v_{G})]\\
\\(6iv_{G}+2iv_{G}^{3}-\omega L+\omega Lv_{G}^{4})+L(v_{G}+1)^{3}(-2i+\omega L(v_{G}+1))]\widetilde{h}_{R}.\end{array}\label{eq: segnale totale lungo z}\end{equation}

A {}``signal'' can be also defined:

\begin{equation}
\begin{array}{c}
\widetilde{S}(\omega)\equiv\frac{\delta\widetilde{\tau}(\omega)}{L}=(1-v_{G}^{2})\exp[i\omega L(1+v_{G})]+\frac{1}{2\omega L(v_{G}^{2}-1)^{2}}\\
\\{}[\exp[2i\omega L](v_{G}+1)^{3}(-2i+\omega L(v_{G}-1)+2\exp[i\omega L(1+v_{G})]\\
\\(6iv_{G}+2iv_{G}^{3}-\omega L+\omega Lv_{G}^{4})+(v_{G}+1)^{3}(-2i+\omega L(v_{G}+1))]\widetilde{h}_{R}.\end{array}\label{eq: sig}\end{equation}

Then the function \begin{equation}
\begin{array}{c}
\Upsilon_{l}(\omega)\equiv(1-v_{G}^{2})\exp[i\omega L(1+v_{G})]+\frac{1}{2\omega L(v_{G}^{2}-1)^{2}}\\
\\{}[\exp[2i\omega L](v_{G}+1)^{3}(-2i+\omega L(v_{G}-1)+2\exp[i\omega L(1+v_{G})]\\
\\(6iv_{G}+2iv_{G}^{3}-\omega L+\omega Lv_{G}^{4})+(v_{G}+1)^{3}(-2i+\omega L(v_{G}+1))],\end{array}\label{eq: risposta totale lungo z due}\end{equation}

is the response function of an arm of the interferometer located in
the $z$-axis, due to the longitudinal component of the massive gravitational
wave arising from the high order gravity theory and propagating in
the same direction of the axis. This longitudinal response function
is proper to \textbf{every} $f(R)$ theory of gravity and is quite
interesting for $f(R)$ theories of gravity which are not banned by
requirements of Cosmology and Solar System tests \cite{key-24,key-25}.

For $v_{G}\rightarrow1$ it is $\Upsilon_{l}(\omega)\rightarrow0$.
Such a response function has been obtained in \cite{key-19} too,
but with a different kind of analysis.

In figures 1 and 2 are shown the response functions (\ref{eq: risposta totale lungo z due})
for an arm of LISA ($L=5*10^{6}Km$) \cite{key-26,key-27} for $v_{G}=0.1$
(non-relativistic case) and $v_{G}=0.9$ (relativistic case). We see
that in the non-relativistic case the signal is stronger as it could
be expected (for $m\rightarrow0$ we expect$\Upsilon_{l}(\omega)\rightarrow0$). 

\begin{figure}
\includegraphics{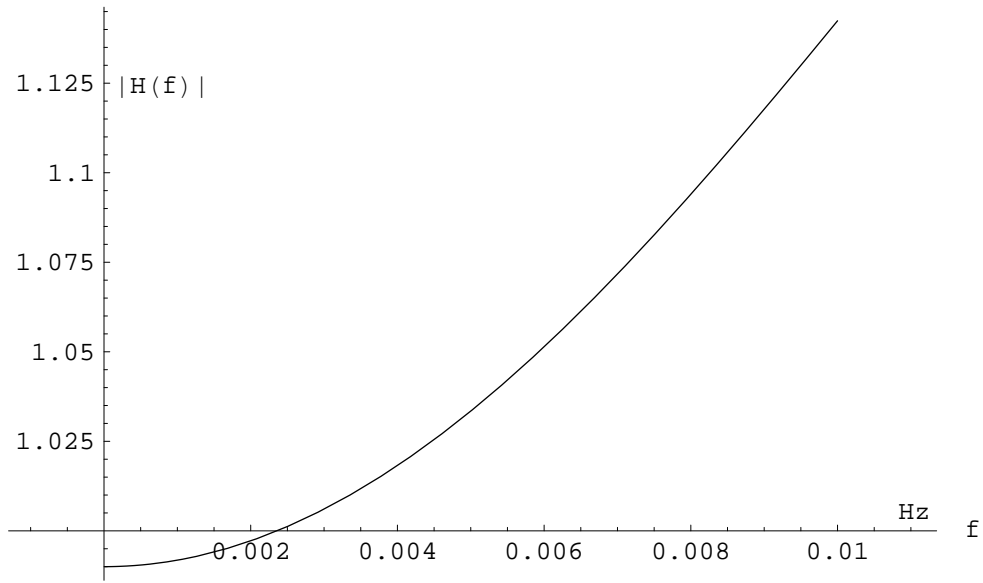}

\caption{the longitudinal response function (\ref{eq: risposta totale lungo z due})
of an arm of LISA for $v_{G}=0.1$ (non-relativistic case) }

\end{figure}

\begin{figure}
\includegraphics{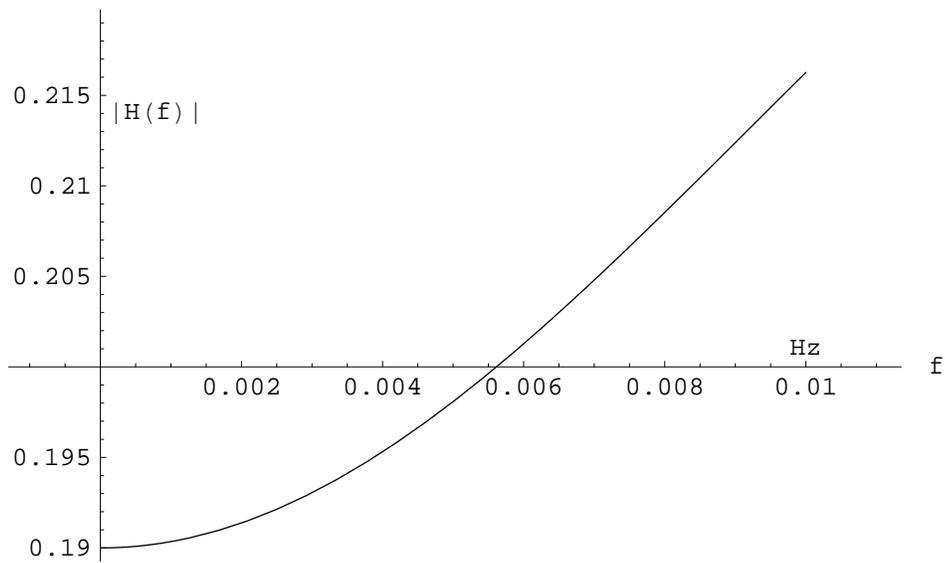}

\caption{the longitudinal response function (\ref{eq: risposta totale lungo z due})
of an arm of LISA for $v_{G}=0.9$ (relativistic case) }

\end{figure}

It is very important to emphasize that, differently from the response
functions of massless gravitational waves, this longitudinal response
function increases with frequency, .i.e , the presence of the mass
prevents signal to drop off the regime in the high-frequency portion
of the sensitivity band. Thus, considering such a high-frequency portion
of the sensitivity band becomes fundamental if LISA would detect massive
GWs arising from $f(R)$ theories of gravity which are not banned
by requirements of Cosmology and Solar System tests \cite{key-24,key-25},
like, for example, the two theories arising from the actions (\ref{eq: high order 1 bis})
and (\ref{eq: high order 1 tris}).

\subsubsection*{Conclusions}

This letter has been a generalization of previous results on gravitational
waves (GWs) from $f(R)$ theories of gravity. In some previous papers,
particular $f(R$) theories have been linearized for the first time
in the literature. Now, the process has been further generalized,
showing that every $f(R$) theory can be linearized producting a third
massive mode of gravitational radiation. In this framework, previous
results are particular cases of the more general problem that has
been discussed in this letter. The potential detectability of such
massive GWs with LISA has been also discussed with the auxilium of
longitudinal response functions.

\subsubsection*{Acknowledgements}

We sincerly thank the anonymous referee for helpful advices which
permitted us to substantially improve the paper.

\end{document}